\documentclass[final,3p,times, authoryear]{article}

\usepackage{geometry}  
\usepackage[T1]{fontenc}  
\usepackage[utf8x]{inputenc}  
\usepackage{lineno}  
\usepackage{fancyhdr}  
\usepackage{enumitem}  
\usepackage{hyperref}  
\usepackage{titling}  
\usepackage{natbib}  
\usepackage{mathtools}  
\usepackage{titlesec}  
\usepackage{lastpage}  

\usepackage[english]{babel}  
\usepackage{amsmath}  
\usepackage{amsfonts}  
\usepackage{amssymb}  
\usepackage{wasysym}  
\usepackage{bbm}  
\usepackage{array}  
\usepackage{xr}  
\usepackage{verbatim} 
\usepackage{float} 

\hypersetup{%
  pdfborderstyle={/S/U/W 1}
}

\setcitestyle{authoryear,round}
\newcommand{\email}[1]{\href{mailto:{#1}}{{#1}}}

\newcommand{\keywords}[1]{\textbf{Keywords}: {#1}}

\newcommand{\wordcount}[2]{\begin{tabular}{rl}%
\textbf{Manuscript word count}: 	& {#1}\\
\textbf{Abstract word count}: 		& {#2}\\
\end{tabular}}

\newcommand{\optincludegraphics}[2][]{}

\newcommand{\optinput}[1]{}

\newtagform{brackets}{[}{]}
\usetagform{brackets}

\newcommand{\thejournal}[1]{Magnetic Resonance in Medicine}

\title{Bilateral breast gradient insert prototype for strong diffusion encoding at 3T}

\immediate\write18{texcount -sum=1,1,1,1,1,1,1 -merge -incbib -dir -utf8 \jobname.tex > \jobname.wcTotal }
\immediate\write18{texcount -sub=none -merge -incbib -dir -utf8 \jobname.tex | grep _main_ | grep -oE '[0-9]+' | python -c "import sys; print(sum(int(l) * w for l, w in zip(sys.stdin, [1, 1, 0, 0, 0, 0, 0])))" > \jobname.wcManuscript }
\immediate\write18{texcount -sub=none -merge -incbib -dir -utf8 \jobname.tex | grep Abstract | grep -oE '[0-9]+' | python -c "import sys; print(sum(int(l) * w for l, w in zip(sys.stdin, [1, 1, 0, 0, 0, 0, 0])))" > \jobname.wcAbstract }

\newcommand{\wcTotal}{\clearpage{\noindent\large{\bf Detailed Word Count} (not to be included for submission)}\verbatiminput{\jobname.wcTotal}}
\newcommand{\wcManuscript}{\input{\jobname.wcManuscript}}
\newcommand{\wcAbstract}{\input{\jobname.wcAbstract}}

\begin{document}

\begin{titlepage}
{\noindent\LARGE\bf \thetitle}

\bigskip

\begin{flushleft}\large
	G.C. Arends\textsuperscript{1,{*}},
	E. Versteeg\textsuperscript{1},
        F. Jia\textsuperscript{2}
        D.W.J. Klomp\textsuperscript{1},
        M. Zaitsev\textsuperscript{2},
        S. Littin\textsuperscript{2,\textdagger},
        C.M.W. Tax\textsuperscript{1,3,\textdagger}
\end{flushleft}

\bigskip

\noindent
\begin{enumerate}[label=\textbf{\arabic*}]
\item Center for Image Sciences, University Medical Center Utrecht, Utrecht, The Netherlands
\item Division of Medical Physics, Department of Diagnostic and Interventional Radiology, Medical Center, University of Freiburg, Faculty of Medicine, University of Freiburg, Freiburg, Germany
\item Cardiff University Brain Research Imaging Centre (CUBRIC), School of Physics and Astronomy, Cardiff University, Cardiff, United Kingdom
\end{enumerate}

\bigskip
\textbf{\textdagger} Shared last authorship
\textbf{*} Corresponding author:

\indent\indent
\begin{tabular}{>{\bfseries}rl}
Name		& Gerrit Cornelis Arends													\\
Department	& Center for Image Sciences													\\
Institute	& University Medical Center Utrecht														\\
Address 	& Heidelberglaan 100														\\
			& 3584 CX														\\
            & The Netherlands														\\
E-mail		& \email{g.c.arends-3@umcutrecht.nl}											\\
\end{tabular}
\vfill

\wordcount{3434}{214}


\end{titlepage}

\pagebreak

\begin{abstract}

\textbf{Purpose:}
Diffusion MRI has shown promise for breast cancer screening, lesion characterization, and treatment response monitoring without contrast agents, but further translation is constraint by the gradient performance of conventional systems. The aim of this work is to develop a single-axis high-performance bilateral plug-and-play breast gradient insert to enable strong-gradient diffusion MRI.

\textbf{Methods:}
An in-house breast gradient insert and bed-tabletop was constructed entirely from commercially available materials, providing a cost-effective solution compatible with existing MRI systems. Its wiring pattern was optimized for torque and force balancing, power dissipation, and target field performance. Evaluation included gradient field characterization, peripheral nerve stimulation simulation verification, and temperature and eddy current assessment. The setup was used for imaging of a diffusion phantom based on soy lecithin across a range of b-values.

\textbf{Results:}
Gradient efficiency reached 2.8\,mT/m/A, enabling local strengths up to 1850\,mT/m (660\,A). No peripheral nerve stimulation was observed during tests on five healthy volunteers. Eddy currents were successfully characterized employed in standard correction methods. Imaging showed the feasibility of b = 10 000 s/mm$^{2}$ acquisitions at TE = 78\,ms versus 161\,ms with scanner gradients.

\textbf{Conclusion:}
This work demonstrates a dedicated bilateral breast gradient insert for safe and feasible strong-gradient breast diffusion MRI, and represents a first step toward dedicated hardware for breast cancer detection and characterization without contrast agents.
\end{abstract}

\bigskip
\pagebreak
\section{Introduction}
Diffusion MRI (dMRI) generates image contrast between tissues with different microstructural properties \citep{Bihan1985, Stejskal1965}, enabling the detection of variations in cellularity, cell shape, orientation, and permeability, amongst others \citep{Guo2002, Sugahara1999DWIglioma, Humphries2007}. By exploiting the sensitivity to these microstructural differences, dMRI can distinguish between normal fibroglandular tissue, benign lesions and malignant lesions in the breast \citep{Guo2002, Ochi2013Diffusion-weightedChanges, Costantini2010DWIBreast}. Consequently, dMRI has emerged as a promising technique for breast cancer screening, demonstrating higher specificity than mammography and dynamic contrast-enhanced MRI \citep{Baltzer2020Diffusion-weightedGroup, Iima2023TheMRI, YAO202336, Clauser2021Diffusion-weightedBiopsy, Hottat2023DWI}. In addition to its diagnostic benefits, dMRI does not require contrast agents, making it a cost-effective and non-invasive alternative to dynamic contrast-enhanced MRI.
However, clinical translation is limited by the performance of conventional whole-body MRI gradient systems, which typically provide maximum gradient strengths of 40–80\,mT/m and slew rates of approximately 200\,T/m/s. These hardware constraints result in prolonged diffusion-encoding times, reduced signal-to-noise ratio (SNR), and limited ability to achieve high b-values and resolutions necessary for accurate breast cancer screening and characterization \citep{Amornsiripanitch2019Diffusion-weightedScreening, Ohlmeyer2021Ultra-HighDetection}.

To overcome these challenges, whole-body strong-gradient MRI systems have been developed and successfully applied in brain \citep{Setsompop2013PushingProject, JONES20188, FAN2022118958} and prostate imaging \citep{Molendowska2024DiffusionGradients, Zhu2024HumanStudy, Bachl2025ProstateDWI}, but their use in breast imaging has thus far not been thoroughly explored. A main drawback of using MRI systems with strong whole-body gradient systems is their limitation by peripheral nerve stimulation (PNS), magnetophosphenes elicitation and cardiac stimulation \citep{Setsompop2013PushingProject, Molendowska2022, klein2021investigating} which significantly reduces the accessible slew rate. 
Other studies have proposed MRI systems with dedicated built-in gradient hardware for a specific anatomical region \citep{Weiger2018ACycle, Huang2021,Feinberg2023Next-generationTesla}. While this approach may become a viable option in the future for large-scale implementation of breast cancer imaging, it currently represents a costly and inflexible strategy for investigating the potential benefits of dedicated strong-gradient dMRI.

The use of plug-and-play gradient inserts provides a unique opportunity to investigate strong-gradient dMRI while not having to dedicate a whole MRI to one specific anatomical region. These gradient inserts can be installed to interface with existing MRI platforms in under one hour, requiring only minimal hardware modifications. Previous studies have demonstrated the feasibility of plug-and-play gradient systems for brain dMRI and functional MRI \citep{Arends2025FeasibilityT, Versteeg2021}, as well as ultrasonic readout techniques \citep{versteegMRM2022}. In a related development, an insertable gradient system was introduced to enable strong-gradient dMRI of the prostate \citep{Zhang2025}. Regarding breast MRI, a single cup breast gradient insert was built \citep{Jia2021DesignBreast, Littin2020BreastGradient}

In this work, we present the experimental realization of an in-house–developed plug-and-play breast gradient insert constructed from commercially available materials and optimized for strong-gradient dMRI based breast cancer screening. The breast-specific design enables higher gradient performance without PNS which is demonstrated through hardware characterization and PNS measurements. Additionally, we showcase the performance gains for dMRI by performing high b-value (b = 10 000\,s/mm$^{2}$) imaging on a diffusion phantom.

\newpage
\section{Methods}

\subsection{Hardware Design and Construction}
\begin{figure}[H]
  \centering
  \includegraphics[width=\textwidth]{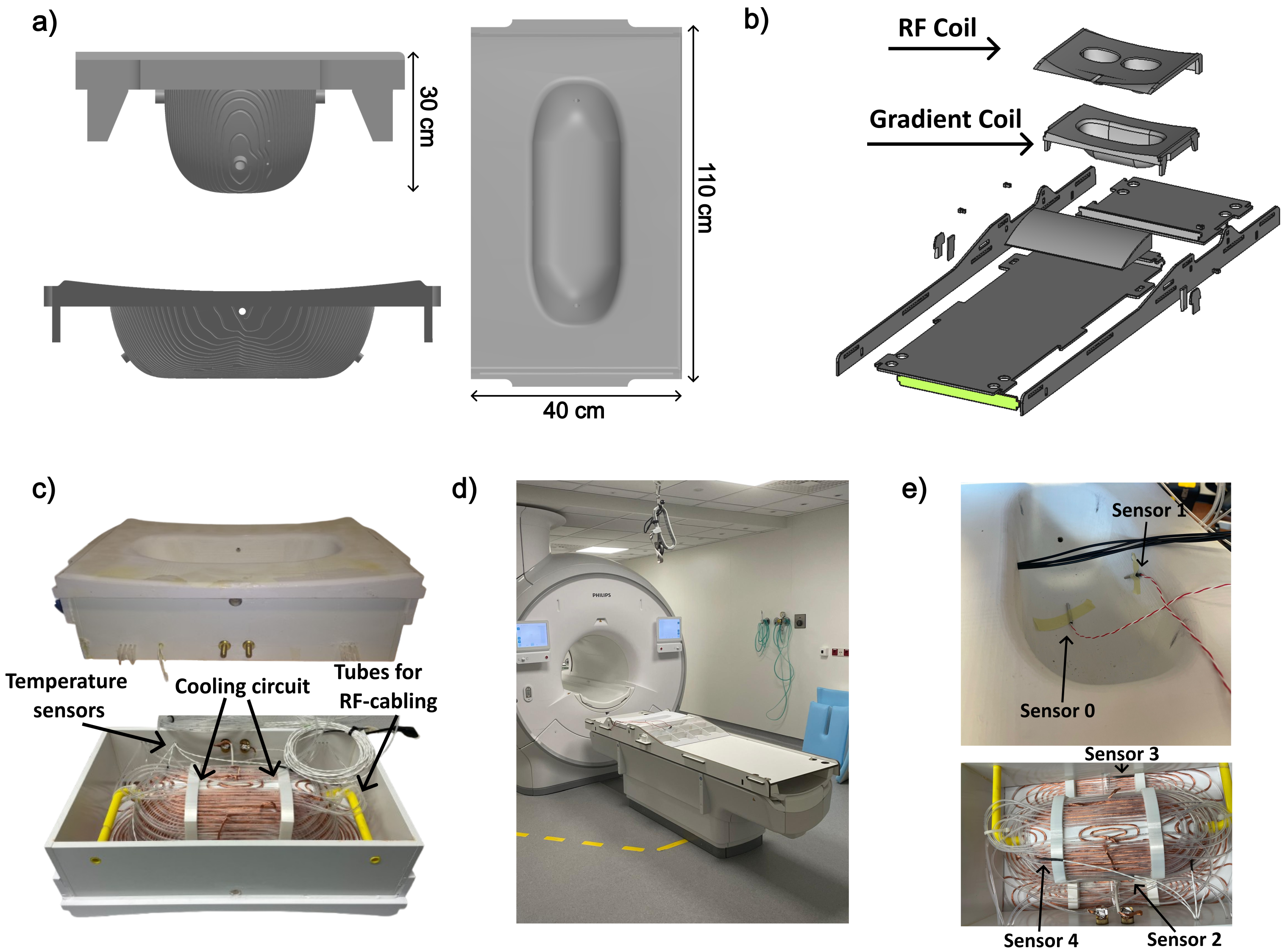}
  \caption{Hardware design \textbf{a)} Top and side views of sketches of the breast gradient with dimensions indicated. \textbf{b)} Exploded view of the CAD model of the complete integrated system. \textbf{c)} Realization of the breast gradient. The top image shows the finalized breast gradient, and the bottom image shows the gradient before being cast into epoxy resin. \textbf{d)} Photograph of the integrated system mounted on the MRI scanner, replacing the standard tabletop. \textbf{e)} Locations of the temperature sensors placed within the breast gradient insert (top image) and embedded in the epoxy near the copper windings (bottom image).}
  \label{fig:1}
\end{figure}

A custom patient bed was constructed from water–jet cut Trespa panels designed in SolidWorks (Dassault Systèmes, Vélizy-Villacoublay, France) (see Fig.\ref{fig:1}b,d). The bed was designed for a wide-bore (70\,cm) 3T MRI system (Philips, Best, The Netherlands) and includes a dedicated compartment for the breast gradient insert. 
The gradient coil surface was designed in Autodesk Inventor (Autodesk Inc., San Rafael, CA, USA) to ensure compatibility with a future radio frequency (RF) coil featuring a cup volume of 1.44\,dm$^{3}$, accommodating approximately 91\% of the female population. Derivation of this estimation was based on the Duke dataset \citep{Saha2018AFeatures} (see supplementary information Fig.S1). The wiring pattern optimization was performed as described in \cite{jia2026designdoublebreastgradient}, aiming to minimize coil power dissipation while preserving local gradient strength. Additional constraints were imposed to ensure a uniform gradient amplitude within coronal slices and to reduce slice-to-slice variation. Furthermore, electromagnetic forces and torques acting on the coil were constrained to maintain mechanical stability. Finite wire thickness was incorporated using a p-norm approach, and CoilGen \citep{Amrein2022CoilGen:Generator} was used in MATLAB (MathWorks Inc., Natick, MA, USA) to convert discrete streamlines into a continuous wire path. A coil carrier with engraved grooves following the wiring path was fabricated via 3D printing (see Fig.\ref{fig:1}a). Round Litz wire with a diameter of 2.8\,mm (105 strands of 0.2\,mm) was glued into the grooves. The Litz wire was subsequently soldered to two brass screw terminals, allowing cables terminated with ring terminals to be connected to the breast gradient using two bolts. Tubes for feeding out RF cables, polyamide cooling tubes and PT-100 temperature sensors were integrated into the assembly, and the completed coil was cast in epoxy resin under vacuum (see Fig.\ref{fig:1}c).

\subsection{Experimental Setup}
A diffusion MRI phantom was constructed using seven tubes containing various soy lecithin concentrations (Carl Roth, Germany). The MRI system body RF transmit/receive coil was used for signal acquisition. The gradient insert was powered with a dedicated amplifier (660\,A/990\,V, Prodrive Technologies) and controlled by an external waveform generator (33500B, Keysight Technologies, Santa Rosa, CA, USA) that interfaced to the MRI using a TCP/IP-connection and TTL-triggering .

\subsection{Safety assessment}
\subsubsection{Temperature Characterization}
Thermal behavior was assessed using embedded PT-100 temperature sensors at hotspots within the winding pattern and on the gradient surface (see Fig.\ref{fig:1}e). Block pulses (50\,ms, 150\,A) were applied every 500\,ms (10\% duty cycle) for 10\,minutes, first without cooling and then with active cooling engaged. This results in a root-mean-squared current of 47.4\,A. The steady-state temperature of the breast cavity and Litz wire was calculated by fitting the measurements to a heating and cooling profile characterization \citep{Chronik2000}.

\subsubsection{Acoustic Characterization}
Acoustic noise measurements were performed for $b=[1000, 2000, 3000, 4000, 5000]\text{ s/mm}^2$ images using a microphone (Behringer ecm8000) placed in the MRI. The audio waveforms were recorded and processed in MATLAB and the setup was calibrated using a sound calibrator (type 4231, Brüel \& Kjær, Denmark). The maximum sound level within each acquisition cycle was extracted and averaged across all cycles to yield the mean peak sound level in dB(A).

\subsubsection{PNS Assessment}
PNS thresholds were measured in five healthy volunteers (two male, three female) positioned prone on the bed with breast gradient. Experiments were performed with subjects lying on a 3D-printed RF coil surface to ensure realistic position with respect to the coil (see Fig.\ref{fig:1}b). Bipolar trapezoidal pulses (50\,ms duration, 3.5\,ms plateau) with rise times of 50, 100, 200, and 400\,$\mu$s were applied four times, after which the volunteer was asked whether they experienced any PNS. Current amplitudes were increased in 50\,A increments up to 600\,A, limited by the gradient amplifier and a maximum slew rate of 2.5\,A/$\mu$s. Rise-time ordering was randomized, and participants were blinded to the conditions. 

Complementary PNS simulations were performed using Sim4Life (ZMT Zurich MedTech AG, Zurich, Switzerland) with the Yoon-Sun female anatomical model. The simulation workflow was as follows. First, the magnetic vector potential generated by the breast gradient coil within the human body model was computed using the Biot–Savart law. This vector potential was subsequently used to calculate the induced electric field throughout the body via a quasi-static electromagnetic solver. Finally, PNS thresholds were estimated using the NEURON solver \citep{hines_neuron_1997}, incorporating modified McIntyre–Richardson–Grill neuronal models \citep{Gaines_Polasek_2018}, based on the computed electric field distribution.

\subsection{Gradient Field Characterization}
The inductance of the coil was measured using an LCR meter (Keysight Technologies, Santa Rosa, CA, USA), enabling estimation of the maximum achievable slew rate. Biot–Savart simulations of the interconnected wire paths were used to compute gradient fields. Experimental data to validate the gradient field simulations were obtained by acquiring phase maps in multiple coronal slices with and without applied gradient blip of 0.5\,ms and 2\,A. Field maps were obtained from the phase differences, and coil sensitivity $\eta$ was calculated as the gradient of the induced magnetic field. Acquisition parameters: FOV (LR, AP) = $320\,\mathrm{mm} \times 320\,\mathrm{mm}$; reconstruction matrix = $80 \times 80$; number of slices = 55; slice thickness = $4\,\mathrm{mm}$; TE = $6.9\,\mathrm{ms}$; TR = $10\,\mathrm{ms}$.

\subsection{Eddy current characterization}
A water bottle phantom was used to estimate induced eddy currents generated by the breast gradient. A 3\,s gradient pulse of $\pm 40$\,A was applied, followed by acquisition of 2D phase images with a TR of 20\,ms for 5\,s after the pulse, resulting in 250 images. To correct for offsets introduced by the gradient amplifier, the mean phase of the images acquired between 2\,s and 5\,s was subtracted. The difference between the $+40$\,A and $-40$\,A acquisitions was then calculated and analyzed using principal component analysis (PCA) to extract the eddy current components, as implemented in \texttt{scikit-learn} \citep{pedregosa2011scikit}. The first principal component was subsequently fitted with a bi-exponential decay function \citep{SPEES2011116} to extract the time constants of the eddy current decay. This procedure was performed in a sagittal and coronal slice. The coronal orientation was repeated three times to improve the signal-to-noise ratio.
Acquisition parameters: FOV = $256\,\mathrm{mm} \times 128\,\mathrm{mm}$; reconstruction matrix = $64 \times 16$; slice thickness = $8\,\mathrm{mm}$; TE = $2.9\,\mathrm{ms}$; TR = $20\,\mathrm{ms}$.

\subsection{dMRI Acquisitions on a soy lecithin phantom}
dMRI scans were conducted on a 3T Philips system. EPI-based diffusion scans were acquired with b-values of [1000, 2000, 3000, 4000, 5000 and 10 000]\,s/mm$^2$ and corresponding TE of [69, 71, 72, 73, 74, 78]\,ms. Sequence optimization for all slices was performed using a nominal sensitivity of $\eta = 2$\,mT/m/A, a slew rate of 1300\,T/m/s and a gradient strength of 300\,mT/m. Slice-wise variation in diffusion encoding was corrected using the simulated coil sensitivity profile via the relationship $b \propto \eta^{2}$. The offset between the MRI scans and the simulations was extracted by fitting the gradient field measurements to the simulations. Reference diffusion scans using the built-in scanner gradients were acquired with TE of [101, 114, 123, 131, 137, 161]\,ms. Acquisition parameters: FOV = $192\,\mathrm{mm} \times 192\,\mathrm{mm}$; reconstruction matrix = $96 \times 96$; slice thickness = $4\,\mathrm{mm}$; number of slices=24; TR = $20\,\mathrm{s}$.\\
ADC estimates from the breast gradient insert acquisition were obtained in two ways: (i) by calculation using the b = 0\,s/mm$^{2}$ acquisition and the slice with $\eta = 2\,\mathrm{mT/m/A}$ from the b = 1000\,s/mm$^{2}$ acquisition, and (ii) by linear least squares (LLS) fitting across multiple slices at different depths of the b = 1000\,s/mm$^{2}$ acquisition normalized by the b = 0\,s/mm$^{2}$ image. 

\newpage
\section{Results}

\subsection{Safety assessment}
\subsubsection{Temperature Characterization}
Temperature measurements showed that the copper winding temperature remained below $70\,^\circ\mathrm{C}$ during 10\% duty cycle operation at 150\,A (see Fig.\ref{fig:2}a). With cooling applied, temperatures decreased to approximately $60\,^\circ\mathrm{C}$ (locations 2 and 3) and $50\,^\circ\mathrm{C}$ (location 4). Temperatures inside the breast-coil cavity remained below $42\,^\circ\mathrm{C}$ without cooling and below $37\,^\circ\mathrm{C}$ with cooling.
Steady-state analysis for sensor 4 predicted $51.5\,^\circ\mathrm{C}$ when cooling was applied. Results on sensor 0, located closer to the patient, showed a steady state temperature of $38.0\,^\circ\mathrm{C}$. Model definition and fit results for the temperature behavior of both sensors are provided in the supplementary information Fig.S2+3.

\subsubsection{Acoustic Characterization}
The mean peak sound levels were [102.8, 102.8, 103.2, 103.5, 101.4] dB(A) for the insert and [101.9, 101.1, 101.7, 101.6, 101.6] dB(A) for acquisitions with b-values $=[1000, 2000, 3000, 4000, 5000]\text{ s/mm}^2$. Analysis of the sound waveforms revealed that the peak sound level is primarily attributed to the EPI readout, with the insert gradient coil showing small elevations of up to 1.9 dB(A) compared to the scanner gradient coil.

\subsubsection{PNS Assessment}
\begin{figure}[H]
  \centering
  \includegraphics[width=\textwidth]{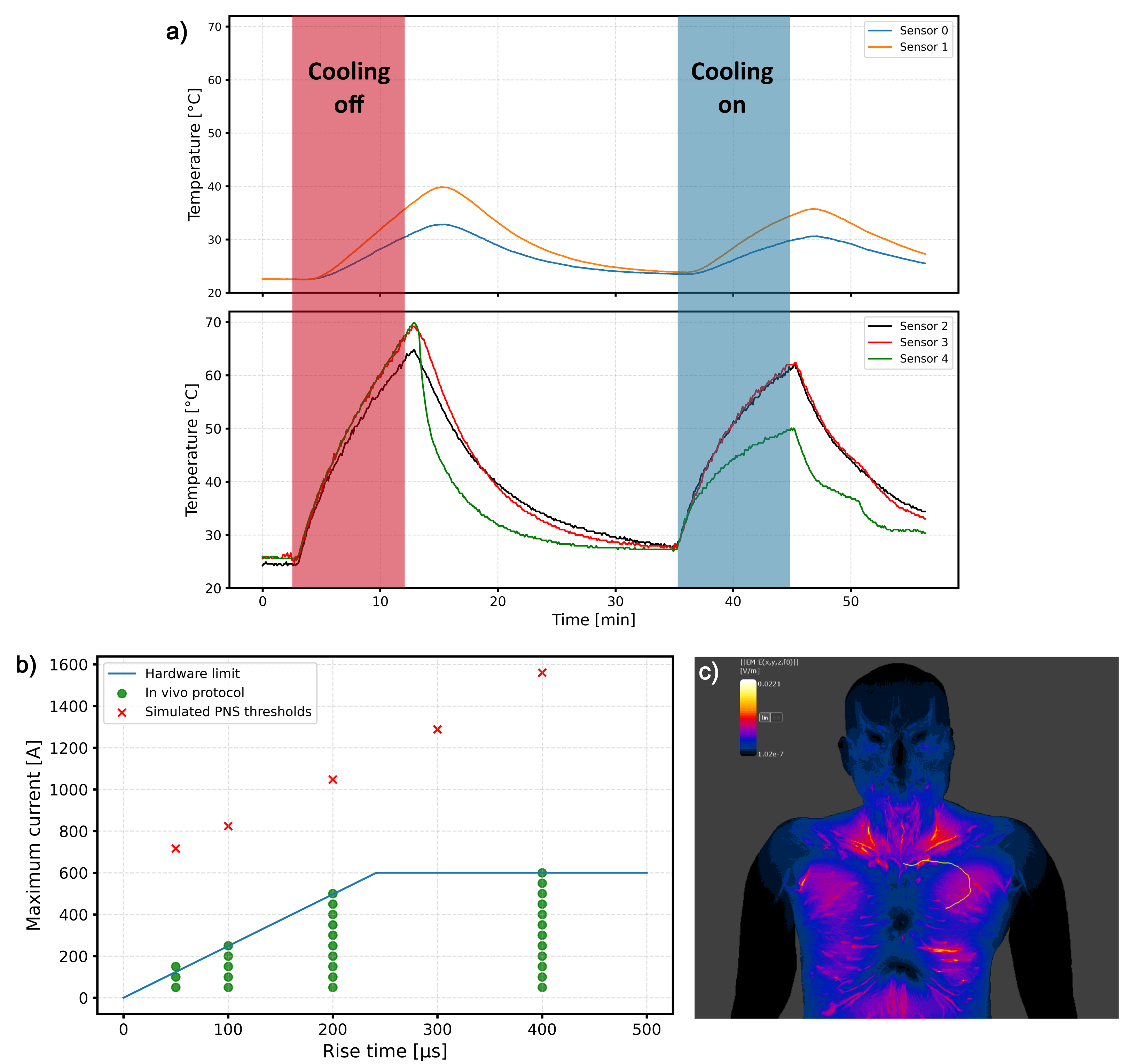}
  \caption{Safety assessment \textbf{a)} Temperature characterization of the breast gradient coil over time. The red-shaded regions indicate periods of current block pulses applied without cooling, while the blue-shaded regions correspond to operation with active cooling. See Fig.\ref{fig:1}e for sensor numbering. \textbf{b)} Hardware limit of the breast gradient, in vivo PNS test protocol and simulated PNS thresholds. No PNS was felt by all volunteers with the indicated in vivo protocol. \textbf{c)} Electric fields (maximum intensity projection) generated by the gradient breast coil for Yoon-sun female model. Electric field within the bones was set to 0. The stimulated nerves were also indicated with the yellow curve.}
  \label{fig:2}
\end{figure}

In vivo PNS tests revealed no stimulation in any of the five volunteers, consistent with simulations predicting PNS onset at 716\,A for a 50\,µs rise time and 1560\,A for a 400\,µs rise time—currents exceeding the tested range (see Fig.\ref{fig:2}b+c). As shown in Fig.\ref{fig:2}b, the applied in vivo protocol remained well below both the hardware-imposed current limits and the simulated PNS thresholds across all investigated rise times, indicating a substantial safety margin. The simulated PNS thresholds exhibited an approximately monotonic increase with rise time, reflecting reduced nerve stimulation for slower gradient switching. Furthermore, the spatial distribution of the induced electric field (Fig.\ref{fig:2}c) demonstrated that peak field intensities were localized primarily in the upper torso regions, consistent with the anatomical proximity to the gradient coil.

\subsection{Gradient Field Characterization}
\begin{figure}[H]
  \centering
  \includegraphics[width=\textwidth]{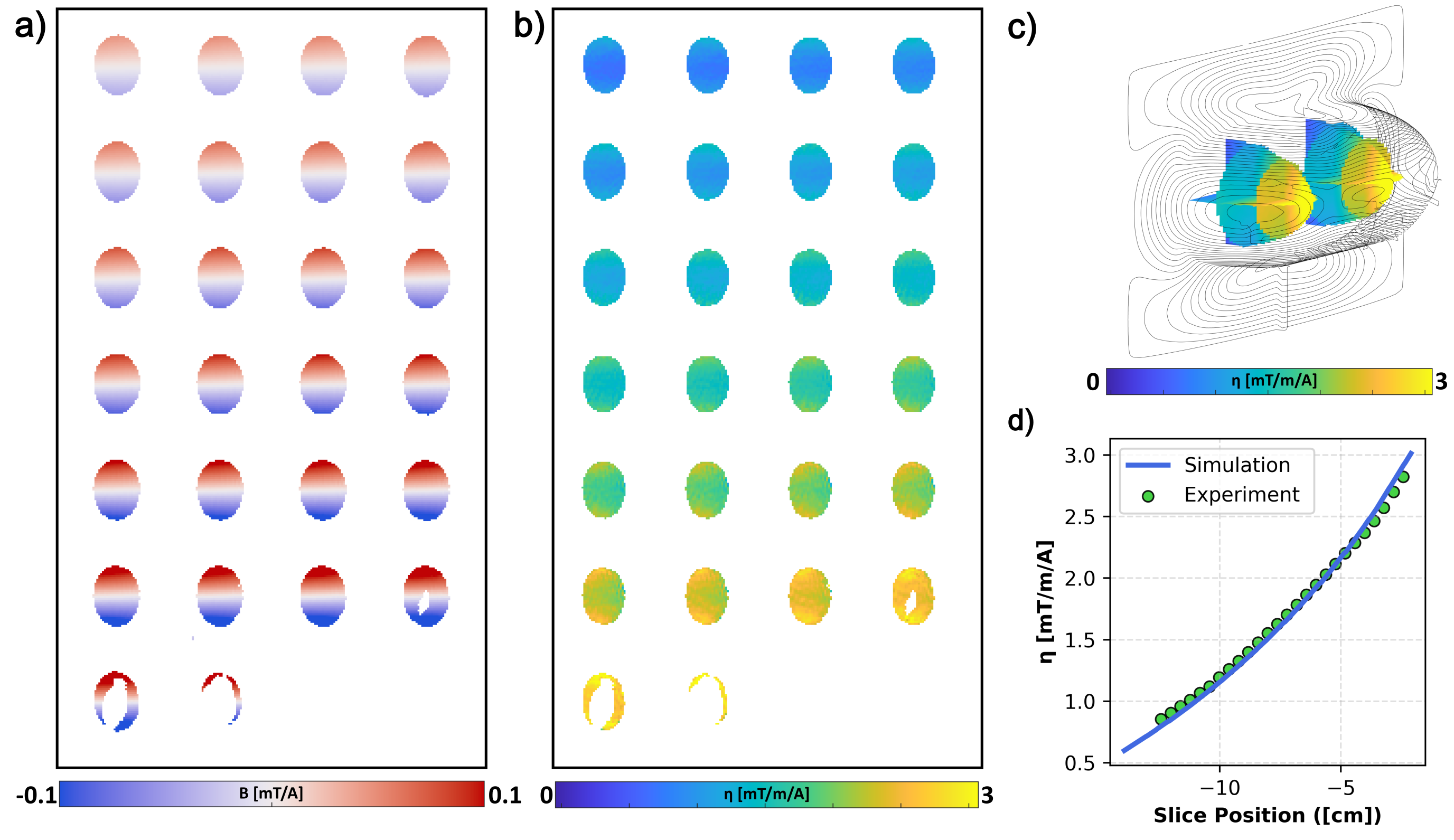}
  \caption{Gradient field characterization \textbf{a)} Difference between phase measurements with and without applied gradient blip. The slice nearest the chest is shown at the top left, while the lowest slice is shown at the bottom right. \textbf{b)} Resulting $\eta$ from the phase measurements in \textbf{a}. \textbf{c)} Simulated $\eta$ with location with respect to the wiring pattern \textbf{d)} Simulated $\eta$ from \textbf{c} plotted versus measurements from \textbf{c} as a function of depth in the breast coil. Slice position 0 is chosen at the wiring pattern.  }
  \label{fig:3}
\end{figure}

The measured magnetic field of the breast gradient insert shows a strong gradient in the z-direction (see Fig.\ref{fig:3}a). The corresponding $\eta$ values range from 0.84\,mT/m/A at the chest wall to 2.8\,mT/m/A, although the last two slices are heavily corrupted by signal loss. The simulated magnetic field gradient calculated from the wiring pattern is visualized in Fig.\ref{fig:3}c. Fig.\ref{fig:3}d shows the measured $\eta$ versus the simulated $\eta$ from Fig.\ref{fig:3}c, demonstrating good agreement between measurements and simulations; however, a small difference in slope is observed.\\
The measured inductance and resistance at 5\,kHz were $129\mu \text{H}$ and $0.271\Omega$ which results theoretically in slew rates between 4200 and 14000\,T/m/s when considering the above mentioned measured $\eta$ and a current of 600\,A and voltage of 800\,V. The full frequency-dependent behavior of the breast gradient is plotted in the supplementary information Fig.S4.

\subsection{Eddy current characterization}
\begin{figure}[H]
  \centering
  \includegraphics[width=\textwidth]{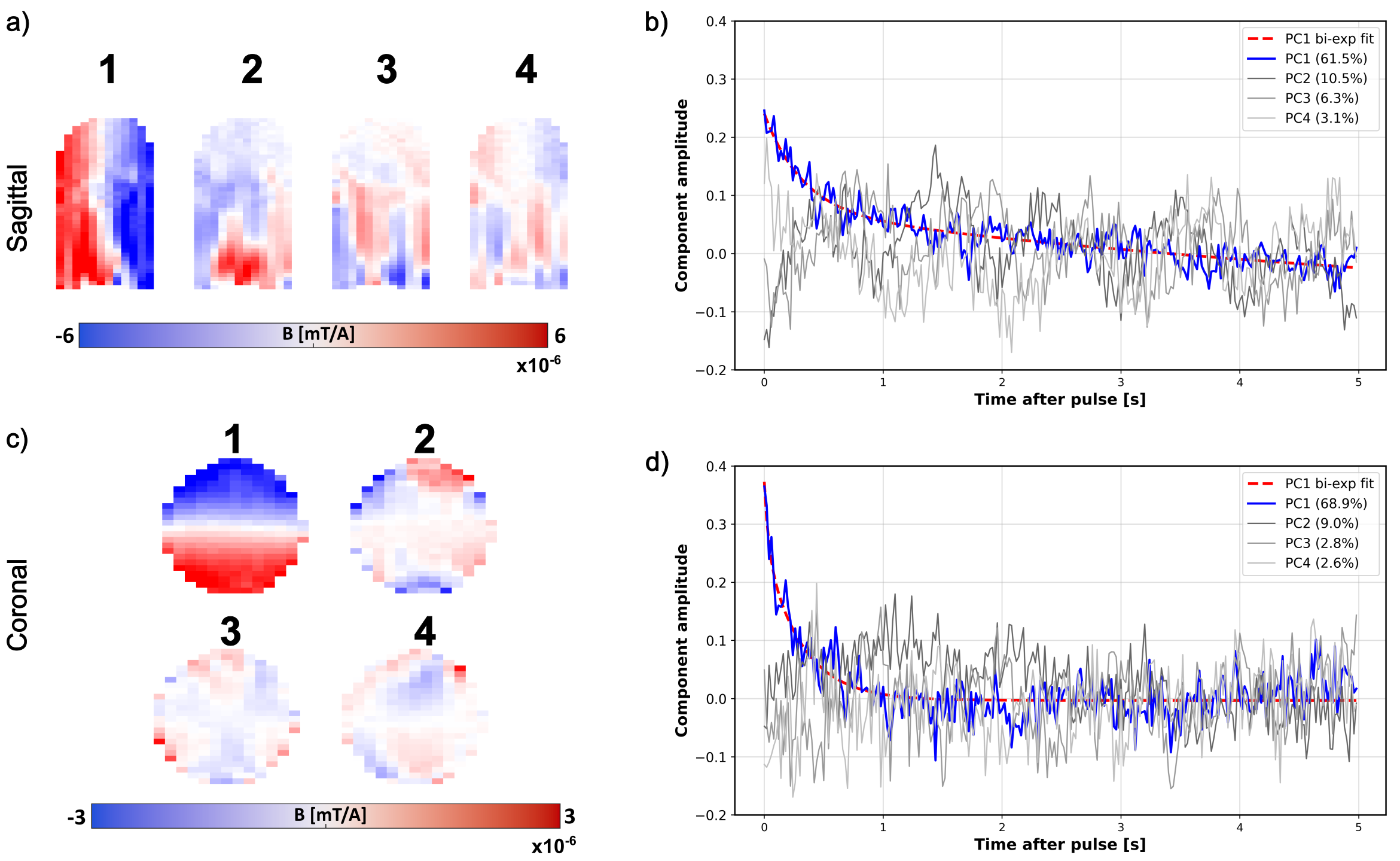}
  \caption{Eddy current characterization \textbf{a)} Sagittal PCA maps corresponding to the first four components. \textbf{b)} Temporal evolution of the first four components corresponding to \textbf{a}. A bi-exponential decay fit is plotted for the first component. \textbf{c)} Coronal PCA maps corresponding to the first four components. Temporal evolution of the first four components corresponding to \textbf{c}. A bi-exponential decay fit is plotted for the first component.}
  \label{fig:4}
\end{figure}

Eddy current characterization of the breast gradient insert is shown in Fig.\ref{fig:4}. Both sagittal and coronal slices of the signal phase acquired after a long diffusion pulse reveal a first-order eddy current field in the z-direction. PCA shows that this component explains 61.5\% of the phase variance in the sagittal measurement and 68.9\% in the coronal measurement. All remaining components account for 10.5\% or less each. A bi-exponential decay fit applied to the first principal component yielded, for the sagittal measurement, a fast decay time constant of 0.32\,s and a slow decay time constant of 7.6\,s. For the coronal measurement, the corresponding fast and slow decay time constants were 0.034\,s and 0.31\,s, respectively. When projecting the first-order eddy current onto the time point corresponding to the end of the diffusion pulse, the resulting gradient is $5 \times 10^{-4}$\,mT/m/A, which corresponds to approximately $0.03\%$ of the main gradient field.

\subsection{dMRI Acquisitions on a soy lecithin phantom}
\begin{figure}[H]
  \centering
  \includegraphics[width=\textwidth]{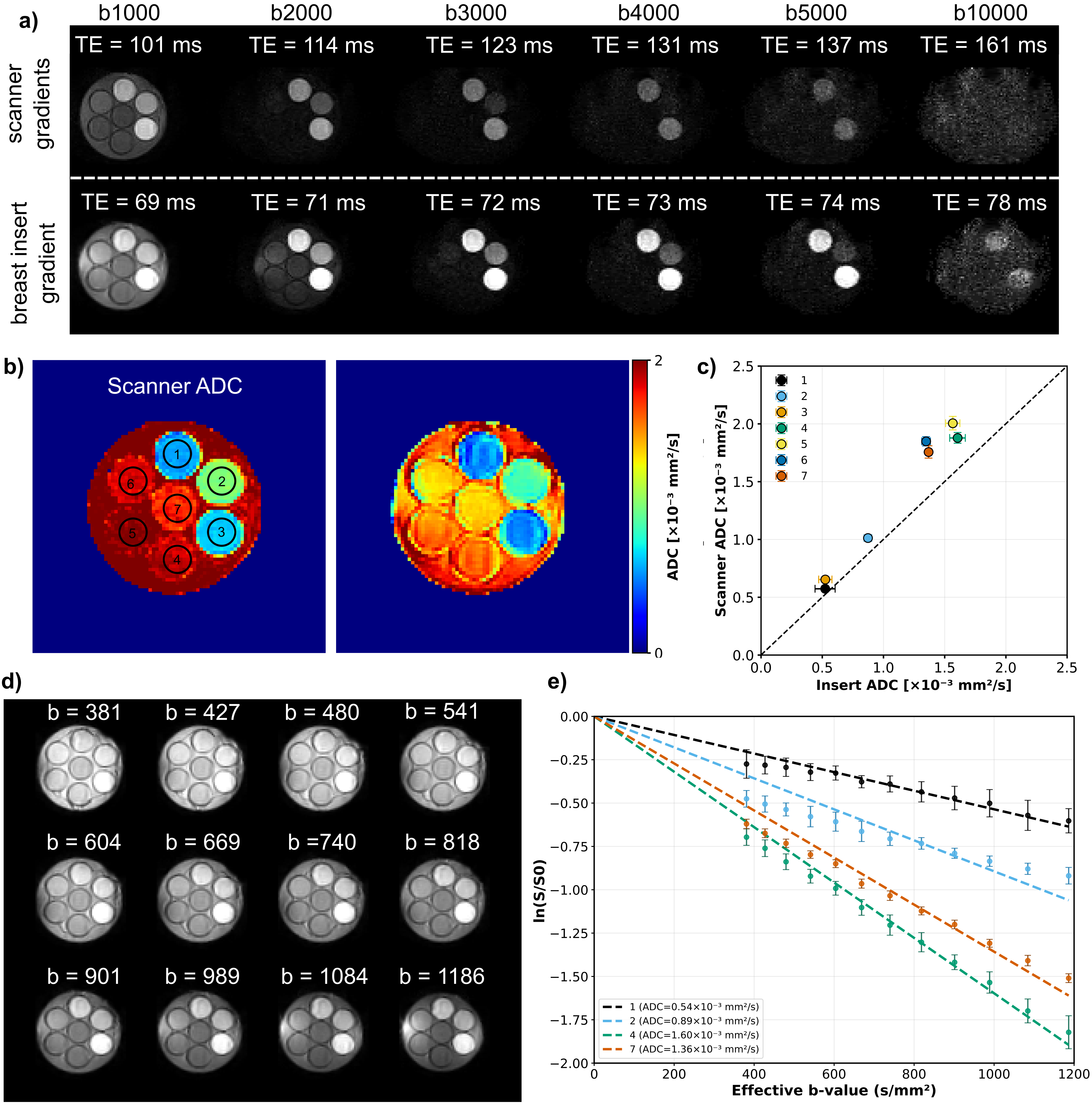}
  \caption{dMRI Acquisitions on a soy lecithin Phantom  \textbf{a)} dMRI images of the diffusion phantom for varying b-values, acquired with the scanner gradients and the breast gradient insert. The corresponding TE is indicated above each plot. \textbf{b)} ADC maps calculated from the b = 1000\,s/mm$^{2}$ acquisition for both the scanner and breast gradient insert acquisitions. Tube numbering is included in the scanner ADC image.  \textbf{c)} Scanner ADC values versus breast gradient insert ADC values for each tube. The standard deviation within the region of interest (see circles in \textbf{b}) is shown in green.  \textbf{d)} Images at varying depths within the breast gradient insert for the b = 1000\,s/mm$^{2}$ acquisition. The corresponding b-value is indicated above each image.\textbf{e)} LLS fits for four tubes of the phantom using slices acquired at varying depths within the breast gradient insert. Legend gives tube number from \textbf{b} and ADC fit result.  }
  \label{fig:5}
\end{figure}

Fig.\ref{fig:5} showcases the dMRI acquisitions of the soy lecithin phantom. Fig.\ref{fig:5}a shows dMRI scans for b-values up to b = 10 000\,s/mm$^{2}$. For these specific acquisition parameters, the breast gradient insert reduced the TE by 32\,ms for the b=1000\,s/mm$^{2}$ acquisitions, and by 83\,ms for the b = 10 000\,s/mm$^{2}$ acquisition. Consequently, the SNR is higher in the breast gradient insert acquisitions compared to the scanner acquisitions. Fig.\ref{fig:5}b+c demonstrate that the ADC estimated using the breast gradient insert is comparable to the ADC derived from acquisitions with the scanner gradients. However, at higher b-values, the ADC obtained with the breast gradient insert shows slightly lower values. Fitting the ADC of four tubes using LLS across multiple slices of the b = 1000\,s/mm$^{2}$ acquisition (see Fig.\ref{fig:5}d+e) yields values consistent with those derived from the single slice shown in Fig.\ref{fig:5}b. Deviations from the fit are seen from all tubes at lower b-values, corresponding to the upper region of the phantom. 

\newpage
\section{Discussion}
\subsection{Implications of the results}
The results demonstrate the successful development and implementation of a prototype breast gradient insert for strong diffusion encoding. The plug-and-play design of the gradient insert, combined with its low cost and relatively straightforward production, enhances the accessibility of strong gradient systems for MRI research. The realized decrease in accessible TE and diffusion times for a given diffusion weighting can beneficially impact breast dMRI by increasing SNR and the range of accessible acquisition parameter settings for microstructural characterization \citep{Jones2018}. Although comparison with acquisitions performed using the scanner gradients demonstrates only a modest increase in signal in the phantom experiments, a larger effect is expected for in vivo breast imaging: Due to the overall lower signal intensity and shorter $T_2$ of breast tissue ($\sim$54\,ms) \citep{Rakow-Penner2006} compared to the phantom (133-792\,ms) \citep{Fritz2023SoyTissues}, the reduced TE is anticipated to have a more pronounced effect on the measured signal intensity.

Safety characterization confirmed that the insert can be operated in an MRI system, considering thermal behavior, PNS, and acoustic noise, when using currents up to 150\,A. Thermal tests showed that the Litz wire temperature stayed below the epoxy melting point (70 °C) during the tests, and the breast gradient insert surface remained under 42 °C, meeting IEC 60601-2-33 touch temperature limits. The steady-state temperature results for the no-cooling period (105\,°C and 89\,°C) highlight the necessity of active cooling for prolonged operation or higher duty cycles. Thermal testing conditions (TR = 500\,ms, 10\% duty cycle, $I = 150$\,A) correspond to a uniform gradient strength of $\sim$125\,mT/m (using $\eta = 0.84$\,mT/m/A) and an achievable $b$-value of $\sim$8000\,s/mm$^2$ for a diffusion time of 25\,ms—well above typical breast dMRI parameters \citep{Baltzer2020Diffusion-weightedGroup}. Furthermore, in vivo PNS tests validated the simulation results, indicating that a next design iteration of the breast gradient insert capable of sustaining higher currents with the same wiring pattern would not be limited by PNS effects. More female volunteers with varying breast size need to be scanned to further confirm this. The localized field of the breast gradient insert may also help mitigate other gradient-induced physiological effects, such as the elicitation of magnetophosphenes, which is particularly relevant for breast MRI, as these effects occur predominantly when the heart is positioned at the isocenter of the MRI system \citep{Molendowska2022}.

Gradient field characterization showed that, at 150\,A, the insert achieves a maximum gradient strength of 420\,mT/m and a mean gradient strength of 250\,mT/m, well above conventional clinical MRI systems (40-80\,mT/m). For uniform diffusion weighted across the whole imaging volume, the achievable gradient is limited by the lowest efficiency ($\eta = 0.84$\,mT/m/A), giving 125\,mT/m at 150\,A, still (3-1.5) times higher than conventional clinical MRI systems.
 
Eddy current characterization revealed a first-order eddy current in the z-direction, closely resembling the gradient field generated by the breast gradient insert. Consequently, eddy current compensation using a gradient impulse response function \citep{Duyn1998TrajectoryCorrection,Addy2012GradientCharacterization,Vannesjo2013GradientCamera}, measured with a field camera or derived from MR acquisitions, could be implemented. 

Finally, the results shown in Fig.\ref{fig:5}c indicate an underestimation of the ADC obtained with the breast gradient insert compared to values derived from the scanner gradients. This discrepancy is most likely caused by suboptimal gradient calibration. Unlike the system gradients, the insert gradient remains to be comprehensively calibrated, potentially resulting in lower effective diffusion weighting than nominally expected. The increase in SNR of the gradient insert acquisitions can also be partially explained by this. Implementation of a gradient impulse response function could mitigate these inaccuracies. 


\subsection{Future work}
Future work will focus on implementing the breast gradient in vivo to show the effect of the reduced TE under realistic physiological conditions. For in-human dMRI studies systematic testing of subject evacuation procedures, assessment of subject comfort during scanning, and additional PNS measurements in a larger cohort of volunteers will be established. To date, no findings have emerged that would constitute a fundamental safety concern for in vivo implementation of the breast gradient.

With its high gradient strength and slew rate, this coil opens possibilities for advanced imaging techniques, such as time-dependent dMRI at high frequencies using oscillating gradients. Numerous studies have investigated the potential of time-dependent dMRI-derived model parameters, such as IMPULSED and POMACE \citep{Reynaud_2017}, as biomarkers for breast cancer subtype diagnosis \citep{Ba2023Diffusion-timeCancer, Wu2025MRCytometry, Li2025Time-dependentMRI}. However, these studies were limited by the performance constraints of conventional clinical scanner hardware or were performed in preclinical systems \citep{Someya2022, Xu2020MRIcellsize}. The use of this dedicated breast gradient insert could ameliorate these limitations. To our knowledge, frequencies up to 50 Hz with b-value 500$\text{ s/mm}^2$ have been reported in these studies \citep{Ba2023Diffusion-timeCancer}, whereas the breast gradient insert enables frequencies of up to 100 Hz (mean gradient strength) and 80 Hz (uniform gradient strength).

For in vivo imaging, a new design based on the presented prototype is currently being implemented by an industrial partner. This coil \citep{Littin2025BreastGradient} is based on the same basic wire layout with only small modifications for manufacturing. Directly cooled hollow copper conductors are being used which enables for higher duty cycles and currents. Assuming 660\,A, this design could reach 1850\,mT/m maximum and 1100\,mT/m mean, surpassing strong whole-body systems (300\,mT/m) \citep{Setsompop2013} while remaining below unilateral breast gradients (3\,T/m maximum) \citep{Littin2020BreastGradient}. Simultaneously, a dedicated RF coil will be designed, mitigating the RF limitations induced by the compact conductive structure of the breast gradient insert. The main focus will be to investigate the effect of the breast gradient insert on the contrast between tumor tissue and benign lesions in raw dMRI images (rather than ADC maps), as higher contrast at elevated b-values has previously been demonstrated there \citep{Molendowska2024DiffusionGradients}. In addition, we will assess its impact on detailed microstructural characterization, which is critical for effective non-contrast MRI–based breast cancer screening, diagnosis, and treatment response monitoring.

\section{Conclusion}
This work demonstrates the feasibility of diffusion MRI with a dedicated plug-and-play breast gradient insert, evaluating safety, temperature, PNS performance, eddy currents, and diffusion imaging. 
The results of this work enable further development of high-performance gradients dedicated for breast dMRI which could improve breast cancer characterization with diffusion MRI. It also enables the wider adoption of strong dMRI experiments within the research community through a cost-effective and technically straightforward modification of existing MRI systems.

\section{Acknowledgments}
We thank Emma Cooijmans and Rokus Valentijn for their work on designing mechanics related to the bed and gradient carrier.
This project is supported by the Eurostars 3 Project "FEM-SCAN: Fast and Efficient MRI Scanning for breast cancer detection"; Project id 3325, and by the Dutch Research Council (NWO) under a VIDI grant with file number 21299.

\bibliography{ references} 

\newpage

\section*{Supplementary information}
\renewcommand{\thefigure}{S\arabic{figure}}
\def\table{\def\figurename{Table}\figure}
\setcounter{figure}{0}

\begin{figure}[h]
  \centering
  \includegraphics[width=\textwidth]{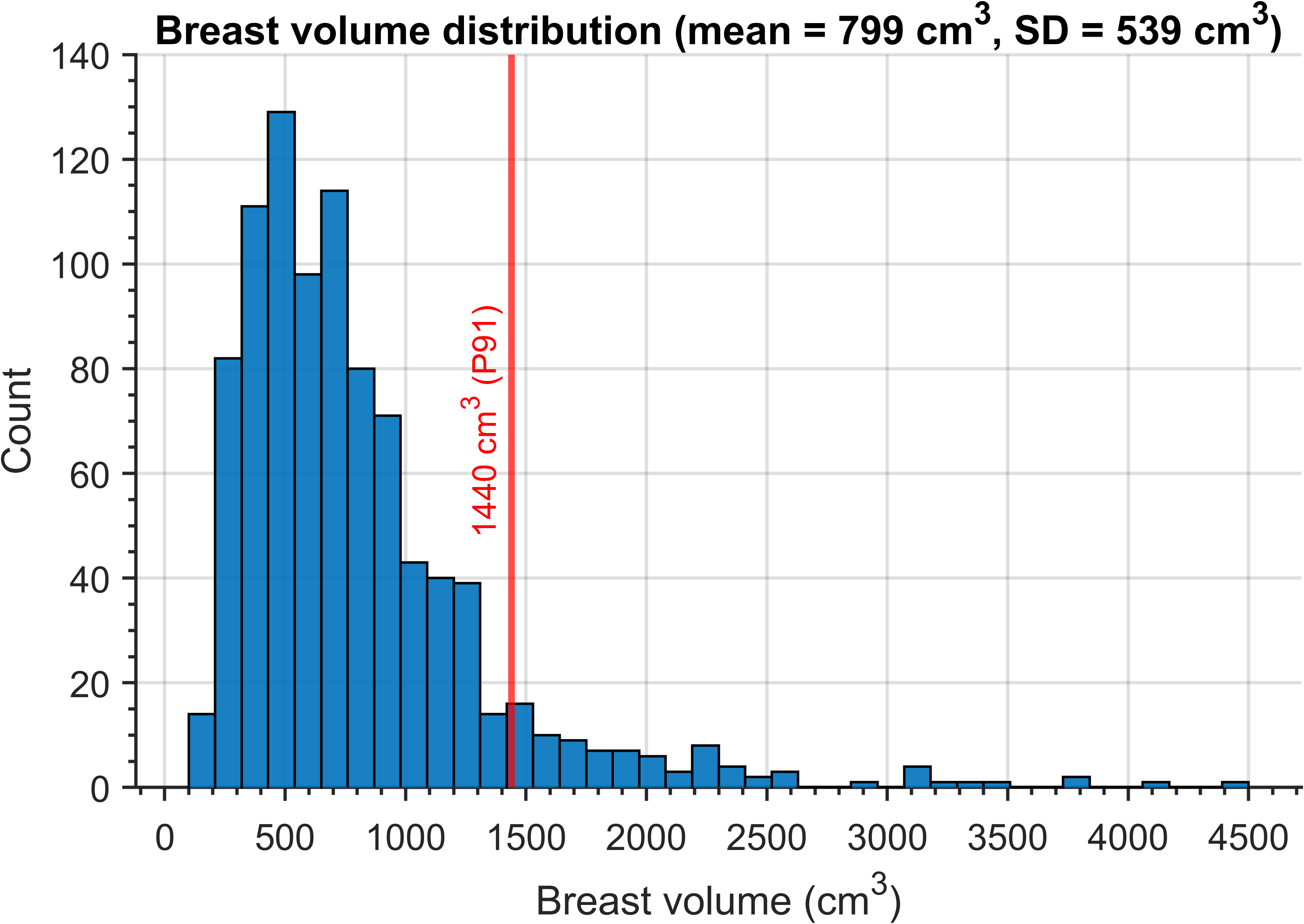}
  \caption{Breast volume distribution of the Duke breast MRI dataset. The total segmented breast volume (sum of both breasts) was divided by two to estimate the mean single-breast volume. The 91st percentile, which was used as the design cutoff for the breast gradient, is indicated.}
  \label{fig:S1}
\end{figure}

\begin{figure}[h]
  \centering
  \includegraphics[width=\textwidth]{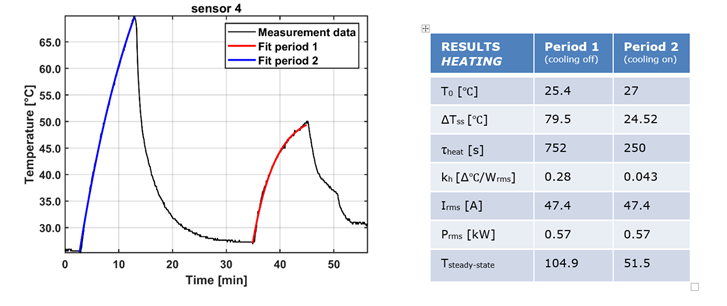}
  \caption{Heating and cooling performance fitting of sensor 4 (integrated in the epoxy of the breast gradient close to Litz wire hotspot). Fits are indicated by red and blue. Fit parameters are shown in the table on the right. }
  \label{fig:S3}
\end{figure}

\begin{figure}[h]
  \centering
  \includegraphics[width=\textwidth]{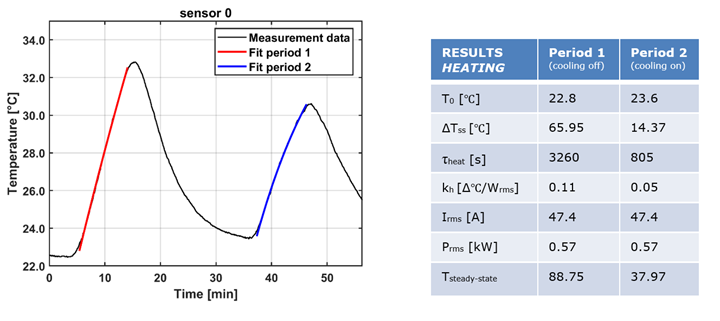}
  \caption{Heating and cooling performance fitting of sensor 0 (located at the bottom of the breast cavity). Fits are indicated by red and blue. Fit parameters are shown in the table on the right. }
  \label{fig:S4}
\end{figure}

\begin{figure}[h]
  \centering
  \includegraphics[width=\textwidth]{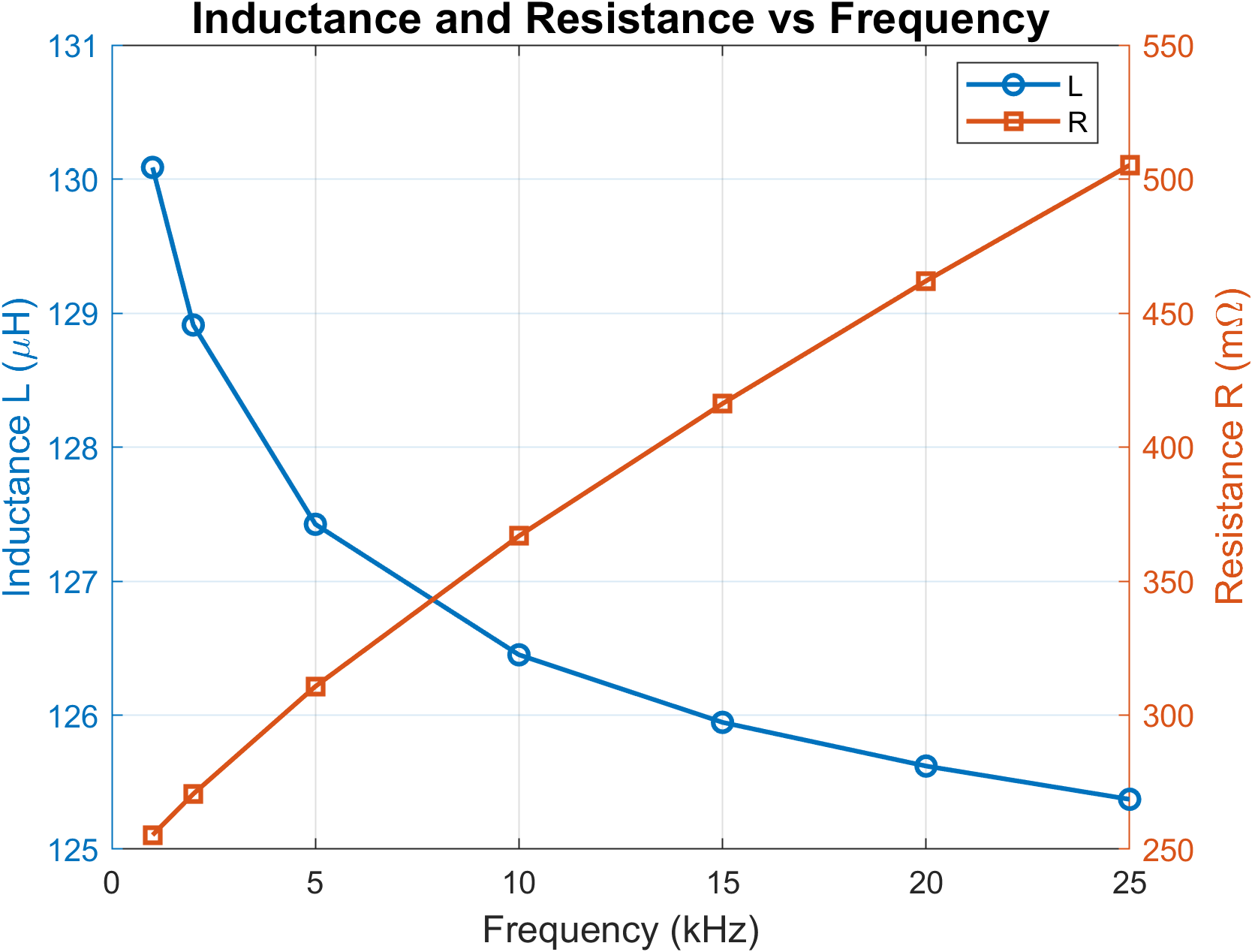}
  \caption{Inductance and resistance of the breast gradient insert versus frequency.}
  \label{fig:S5}
\end{figure}

\end{document}